\documentclass{elsart}
\usepackage[dvips]{graphicx}
\usepackage{amsmath}

\begin{document}
\title{Example of a possible interpretation of  Tsallis entropy}
\author{G. Wilk$^{a}$, Z.W\l odarczyk$^{b,c}$}
\address{$^{a}$The Andrzej So\l tan Institute for Nuclear Studies,
                Ho\.za 69; 00-689 Warsaw, Poland\\ e-mail: wilk@fuw.edu.pl\\
         $^{b}$Institute of Physics, \'Swi\c{e}tokrzyska Academy,
               \'Swi\c{e}tokrzyska 15; 25-406 Kielce, Poland;\\ e-mail:
               wlod@pu.kielce.pl\\
         $^{c}$University of Arts and Sciences (WSU), Weso\l a 52,
                  25-353 Kielce, Poland\\
                  }

 {\today}

{\scriptsize Abstract: We demonstrate and discuss the process of
gaining information and show an example in which some specific way
of gaining information about an object results in the Tsallis form
of entropy rather than in the Shannon one.

\noindent {\it PACS:}  05.20.-y; 05.70.-a; 05.70.Ce

\noindent {\it Keywords}: Nonextensive statistics, Correlations,
Fluctuations}

Some time ago the notion of information in the form dictated by
Shannon entropy \cite{S} was used by us to study multiparticle
production processes \cite{WW}. Later we found that its Tsallis
version \cite{T} is more suitable \cite{NUWW}, because the
nonextensivity parameter $q$ characterizing Tsallis entropy bears
information about the intrinsic fluctuations in the physical
system \cite{qWW}\footnote{Actually, what we later call entropy
$S_q=H$, was introduced independently before, see for example
\cite{E1} and then rediscovered again by Tsallis in
thermodynamics.}. The finding that $q$ can be interpreted as a
measure of such fluctuations resulted in a new subject called {\it
superstatistics} dealing with all kinds of fluctuation, see
\cite{SuperS}. In all these studies we used either the Shannon or
Tsallis  form of information entropy without, however, any deeper
understanding or argumentation about what in fact makes them so
different, i.e., without deeper thoughts about the details of the
process of gaining information leading to one or another form of
information entropy.

In this note we shall concentrate on this problem, using as our
basic tool a widely known example  presented in \cite{K}) which
demonstrates how to deduce in a simple manner the form of Shannon
informational entropy by considering the process of finding the
location of some object in a prescribed phase space (like, for
example, a point on a sheet of paper). We shall develop an
equivalent procedure resulting in Tsallis entropy instead. In
particular we shall demonstrate, using these examples, how the way
in which one collects information about an object decides the form
of the corresponding information entropy. As already mentioned
above we shall concentrate only on the comparison between Shannon
\cite{S} ($S=S_{q=1}$) and Tsallis \cite{T} ($S_q = S_{q\ne 1}$)
forms of entropy:
\begin{equation}
S = - \sum_i p_i \ln p_i \qquad \Leftrightarrow \qquad S_q = -
\sum_i p_i \ln_q p_i = - \sum_i p_i \frac{ \left ( 1 -
p^{q-1}_i\right)}{1-q} .\label{eq:ST}
\end{equation}

Acting in the same spirit as in \cite{K}, consider a system of
size $V_0$ and divide it into cells of size $V$ each; we then have
$M=V_0/V$ such cells (divisions). Suppose now that in one of these
cells an object is hidden (we shall call it a {\it particle} in
what follows) and that the probability to find it in a cell is the
same for all cells and equal $\frac{1}{M}$. The corresponding
Shannon entropy, describing the situation of finding this particle
in one of the cells, is:
\begin{equation}
S = - \sum_{i=1}^{M} \frac{1}{M}\ln \left(\frac{1}{M}\right) .
\label{eq:Shannon}
\end{equation}
Suppose that the cells were formed by consecutively dividing
previous cells into two equal parts and that we have performed $f$
such divisions. Then $M=2^f$ and Shannon entropy
(\ref{eq:Shannon}) corresponds conventionally to $S= \log_2 M = f$
bits of information. The tacit assumption is that to locate a
particle in the system is equivalent to finding the respective
cell containing this particle\footnote{~Actually in \cite{K} this
particle was supposed to be pointlike and one attempted to find
its location inside some system. To do this, this system was
consecutively divided in halfs with {\it a  priori} equal
probabilities to find this particle (point) in one of the two
cells and this procedure was continuing until the desired accuracy
was obtained. In our case this accuracy dictates the number of
cells into which our system is divided. }. In such an approach,
entropy equals just the number of YES/NO questions needed to
locate the selected particle. In a sense, our particle is {\it
structureless}, i.e., it has no additional features which would
have to be investigated before its proper recognition; the
localization of the particle is therefore equivalent with its
recognition.

Suppose, however, that the particles we are searching for have
some additional features one has to account for and that in a cell
there can be more (or less) than one particle. It is obvious that
in such a case the localization of the cell as performed above is
not equivalent to the recognition of the right particle itself.
One can now be faced with two situations:
\begin{itemize}
\item One finds the cell with a particle in it but one is still
not sure that this is the right ("true") particle; some additional
search involving additional features mentioned above is required -
one needs more information than in the usual case.

\item One recognizes the right particle already before the search
of the proper cell is finished, it means that some information
offered is redundant - one needs less information than in the
usual case.
\end{itemize}
The problem now is: how to quantify this problem? There is {\it a
priori} an enormous number of factors which result in those
additional features which should apparently be accounted for. On
the other hand, from our point of view, all of them are, in a
sense, identical because they simply transform the originally {\it
structureless} particle to a particle endowed with some {\it
structure} which can vary from one particle to the other. Let us
therefore concentrate on a simplest possibility and assume that it
is enough to replace the single particle considered originally in
\cite{K} by a number of identical particles endowed with some
artificial {\it size} $\nu $, which can occupy each cell.
Identical means therefore that all particles have the same size
which they keep all the time. Notice that:
\begin{itemize}
\item in the whole volume $V_0$ considered one can only put
$N=V_0/\nu$ particles;

\item in a given cell one can only put $k=V/\nu$ particles; it is
very important to realize that one can have $k > 1$ as well as $k
<1$ (in addition to the original case corresponding to $k=1$).

\end{itemize}

\begin{figure}[t]
  \begin{center}
   \includegraphics[width=11.0cm]{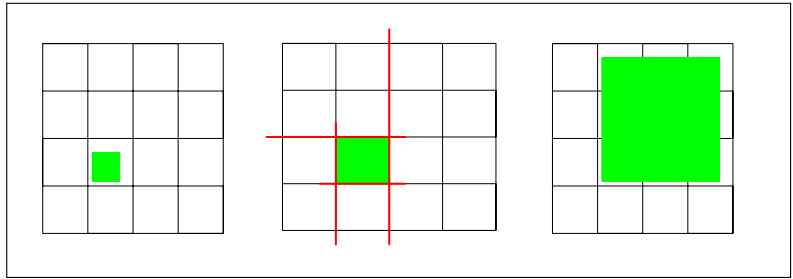}
   \caption{Schematic illustration of three possible situation
            encountered when gathering information. Left panel: whatever number of divisions
            the object we are looking for is {\it smaller} then the actual cell and can
            require some additional (in comparison to the usual) search. Middle panel:
            there is no additional information needed to locate the object. Right panel:
            object {\it shows itself} (and therefore can also be identified) already before
            the usual amount of search (i.e., divisions) is performed. Only the middle panel
            situation leads to the Shannon form of entropy, the remaining two result in the Tsallis
            entropy instead.}
   \label{Figure1}
  \end{center}
\end{figure}

As before we again attempt to locate the selected particle in our
system in a most effective way (i.e., by using only the minimal
possible amount of information). The probability to choose the
cell with this particle is $p=1/M=V/V_0$. However, now the size of
the particle matters and the cell can be occupied by a number of
particles among which we must choose the one we are looking for.
As illustrated in Fig. \ref{Figure1} one can encounter three
typical situations:
\begin{itemize}
\item Even when one finds the right cell one still has to search
for a while before deciding that the chosen particle is the right
one. In our example this is visualized by the fact that the
particle is {\it smaller} then the cell and there can be more than
one particle per cell (notice that {\it more} does not mean here
that the actual number is an integer, it can be any positive
number), cf. left panel of Fig. \ref{Figure1}.

\item It can happen that one is sure that the chosen particle is
the right one even before the right cell has been identified. In
our example this corresponds to a situation when the particle is
{\it bigger} than the cell. This means that the particle occupies
more than one cell (again, {\it more} means that this is any real
number greater than unity but smaller than the maximally allowed
number of cells equal $V_0/V$), cf. right panel of Fig.
\ref{Figure1}.

\item The information needed to locate the particle is the same as
to find the right cell. In our example it simply means that
volumes of cell and particle are equal and there can be only one
particle per cell., cf. the middle panel of Fig. \ref{Figure1}.
\end{itemize}
To continue along this line let us notice that one (chosen)
particle can register in a cell with probability $p$ or not
register in this cell with probability $1-p$. In the case when
this particle is not identified with the cell (as is the case in
the case of counting the Shannon entropy), in this cell there can
be maximally $k-1$ other (false) particles. Assuming independence
of events, the probability of the occurrence $r$ such particles in
the cell is $p^r$. Therefore, on the average, the probability to
register a false particle (and not the chosen one) per one false
particle is\footnote{~~~~~Here and below we are formally using the
symbol of summation in spite of the fact that, as stated before,
$k=V/\nu$ is not necessary an integer. Therefore, when necessary,
one has to use continuous $k$ and replace summation by a suitable
integration to calculate the corresponding $\langle \dots \rangle$
quantities, as for example,
\begin{equation}
                \langle p_{k-1}\rangle = \left\{
                \begin{matrix}
                \frac{-\ln(p)}{k-1}\int^k_1 p^r dr & {\rm for} &
                k>1 \\
                \frac{-\ln (p)}{1-k} \int^1_k p^r dr & {\rm for} & k<1
                \end{matrix}
                \right\}
                = - \frac{p}{k-1} \left( p^{k-1} - 1\right)
                \quad \stackrel{k\rightarrow
                1}{\Longrightarrow}\quad -p\ln(p)
                .\nonumber
                \end{equation}
One gets the same limiting behavior directly applying the
$k\rightarrow 1$ limit to the integral.
                }
\begin{equation}
\langle p_{k-1} \rangle = \frac{1-p}{k-1} \sum^{k-1}_{r=1} p^r = -
\frac{p}{k-1}\left(p^{k-1} -1\right) . \label{eq:choosing}
\end{equation}
Notice that by doing so we are also tacitly assuming that any
false choice also removes (or equivalently marks somehow) the
falsely chosen particle as misidentification.

Before proceeding further, a few words of conditions under which
formula (\ref{eq:choosing}) may be valid are in order. Our picture
could physically correspond to the situation in which we perform
measurements with noise ($q>1$) or when errors connected with the
measurement exceed the size of the cell ($q<1$). In this case our
object is identified in a number of cells (in other words, in this
case our cells are not "mutually exclusive" as they were in the
usual situation leading to Shannon entropy). One can also
encounter a situation when the cells are not refined enough in
phase space, equivalent to the case in which cells are exactly
known but the location of our object (particle) is not fixed. All
such situations eventually lead to Eq. (\ref{eq:choosing}).

To continue, the question we have to answer now is: what is the
corresponding entropy in this case, or - what is the analogy to
YES/NO questions in the previous case, where the number of
questions was the entropy? We argue that the analogy  to YES/NO
questions in this case is the sum over all cells of the
probability {\it to not register} the particle (i.e., probability
to register only false particles). Notice that Eq.
(\ref{eq:choosing}) gives us the gain of information we get from a
single cell. For a system of $2$ cells, $i=1,2$, with $p_1=0$ and
$p_2 =1$ one has $H=0$, whereas for $0 < p_i < 1$ one gets $H >
0$. Therefore one can say that the entropy for a system of $M$
cell is given by
\begin{equation}
H = \sum^M_{i=1}\langle p_{k-1}\rangle = -
\sum^M_{i=1}p\frac{p^{k-1}-1}{k-1} ,\label{eq:H}
\end{equation}
which in our case can be rewritten in the following form,
\begin{equation}
H = -
\sum^M_{i=1}\frac{V}{V_0}\frac{\left(\frac{V}{V_0}\right)^{\frac{V}{\nu}-1}-1}{\frac{V}{\nu}
- 1} . \label{eq:Hprim}
\end{equation}
\begin{figure}[t]
\begin{center}
   \includegraphics[width=9.0cm]{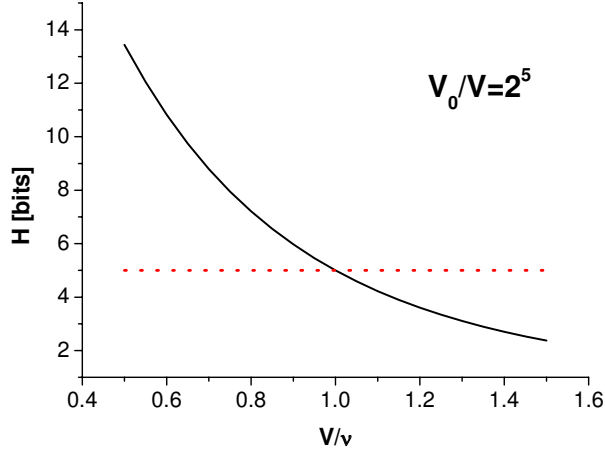}
   \caption{Illustration of the dependence of $H$ on $k=V/\nu$ for the
   number of cells equal to $V_0/V=2^5$ (full line). The Shannon entropy
   (dotted line) for the same number of cells is equal $S=5$ (because we have now $5$ bits).
   Notice that $H > S$ for $k<1$ and $H<S$ for $k>1$.}
   \label{Figure2}
\end{center}
\end{figure}
Notice that entropy defined by Eq. (\ref{eq:H}) or
(\ref{eq:Hprim}) is formally identical in its form with the
Tsallis entropy \cite{T} with $q=k=V/\nu$. Its behavior is
illustrated in Fig. \ref{Figure2}. It has the following
characteristic properties:
\begin{itemize}
\item $H \rightarrow 0$ if $V \rightarrow V_0$, i.e., when one has
just one cell (in this limit there is no information left in the
system, see Fig. \ref{Figure2}, $k>1$ case);

\item When $\nu \rightarrow V_0$, i.e., when the particle we are
looking for covers all cells (becoming therefore identical with
the system itself), the entropy $H$ grows monotonically and
reaches the limiting value
\begin{equation}
H = \frac{M}{M-1}\left( M^{1-1/M} - 1 \right) < M .
\label{eq:limiting}
\end{equation}
It is interesting to note that when $M$ becomes very large then $H
\longrightarrow M$ in this limit. This should be contrasted with
the much milder behavior of Shannon entropy which in this limit
becomes $S \propto \ln M$ (cf. Fig. \ref{Figure2}, $k<1$ case).
The difference arises because in this case in order to fully
identify a particle in the system one must go through {\it all
cells} with positive outcome, not only through a sunset of them as
is done in the procedure leading to Shannon entropy.

\item $H \rightarrow S=-\sum_{i=1}^M\frac{V}{V_0}\ln \left(
\frac{V}{V_0}\right)$ if $\nu \rightarrow V$, i.e., when the
particle we are looking for has the same size as the cell; in this
case $k \rightarrow 1$ and we recover the classical definition of
Shannon entropy as given by Eq. (\ref{eq:Shannon}) (see
intersection for $k=1$ in Fig. \ref{Figure2}).
\end{itemize}

It is interesting to mention at this point that in the case where
we would be interested in finding not one but some number, say
$k_2$, of particles among $k_1$ particles (notice that in the
previous discussion $k_1 = k$ and $k_2=1$) then Eq.
(\ref{eq:choosing}) would be replaced by
\begin{eqnarray}
\langle p_{k_1-k_2}\rangle = \frac{(1-p)^{k_2}}{k_1-k_2}
\sum^{k_1-k_2}_{r=1} p^r =  \frac{1}{k_1-k_2}
p(1-p)^{k_2-1}\left(p^{k_1-k_2}-1\right), \label{eq:choosing1-2}
\end{eqnarray}
which would then result in the two parameter form of entropy $H$,
even more general than the Tsallis entropy (see \cite{MT} for
examples and discussions of such entropies, we shall not pursue
this point further here).

Let us finally consider two systems consisting of $M$ elements
each: $A$ and $B$. Let us proceed in the same way as above,
treating each system independently with probabilities $p_A$ and
$p_B$ replacing $p$. Suppose we are looking for two particles: one
from $A$ and one from $B$ (and let us assume that they have the
same structure in both systems). Notice that even when
individually $p=p_A\cdot p_B$, their average does not factorize
but is given by the following expression:
\begin{eqnarray}
\langle p_{A,k-1}\cdot p_{B,k-1}\rangle &=& \frac{1-p_Ap_B}{k-1}
\sum^{k-1}_{r=1}p_A^rp_B^r = \frac{p_Ap_B}{k-1}\left( 1 -
p_A^{k-1}p_B^{k-1}\right)
=\nonumber\\
&=& p_B\langle p_{A,k-1}\rangle + p_A \langle p_{B,k-1}\rangle +
(1-k)\langle p_{A,k-1}\rangle \langle p_{B,k-1}\rangle .
\label{eq:pApB}
\end{eqnarray}
This in turn means that ($i$ denotes summation over cells in $A$
and $j$ in $B$)
\begin{eqnarray}
H_{A,B} =  \sum^M_i\sum^M_j \langle p_{A,k-1}p_{B,k-1}\rangle =
H_A + H_B + (1-k) H_A H_B ,\label{eq:HAHB}
\end{eqnarray}
i.e., that entropy $H$ is nonadditive.

To summarize: we have demonstrated on a simple example how the way
one gets information on the system leads to different forms of the
information entropy when this is understood as some suitable
measure of this information. The most general form, encompassing
the situations in which the object we are looking for has some
internal degrees of freedom (here summarily described by endowing
it with some artificial {\it size}), is the one described by
entropy $H$ as defined by Eq. (\ref{eq:H}) which has the form of
the Tsallis entropy \cite{T}. The Shannon entropy
(\ref{eq:Shannon}) is (at least in the example studied here) only
a limiting case corresponding to a structureless object. As one
can see in Fig. \ref{Figure2} it corresponds to a single point
only for which $k=q=1$. Otherwise one always gets Tsallis entropy.
One should bear in mind that this is a really very simple (if not
simplistic) analysis, assuming only discrete situations. On the
other hand, we argue that it already explains the essence of the
difference between $H=S_q$ and $S=S_{q=1}$ in Eq. (\ref{eq:ST})
(and it also bears a potential for even further developments as
witnessed by Eq. (\ref{eq:pApB})).

One must, however, bear in mind the possible limitation of our
approach caused by our particular inclination towards problems of
high energy multiparticle production processes where intrinsic
fluctuations are very important in a proper description of systems
in which some given initial amount of energy is converted into
finally observed particles (hadrons) in the process called {\it
hadronization}. This link is behind our, probably peculiar view of
entropy and its connection with some physical processes. As far as
we can tell, the classical works on entropies and their properties
(like, for example, \cite{E1}, see also the most recent review
\cite{E2}) are rather mathematical in their form and scope, and,
for example, not directly applicable to the subject mentioned
above. One should mention at this point previous attempts to
extend Shannon entropy in which either nonadditivity of the
entropy measure was important, not the additional one parameter
\cite{Y}, or in which a two-parameter family of trace formula
entropies were discussed \cite{KLS}.

The natural question coming to mind is about a possible
application of the method proposed here to other types of entropy.
Because of the enormous number of possible entropies (see, for
example, the list in \cite{EE} and in \cite{E2}), this goes
outside the limited goal of our work. Nevertheless, closing our
presentation we make a few comments concerning the widely used
Renyi entropy. It also has an extra parameter (often denoted by
$q$) , however, contrary to Tsallis entropy it is extensive. The
meaning of the $q$ parameter used in Renyi entropy is quite
different from that in Tsallis entropy. By construction, as
discussed in detail in \cite{MW},  Renyi entropy $R_q$ is
sensitive to non-uniformity of the measure of the phase space with
$q$ being a kind of control parameter specifying the regions of
phase space of interest. From the point of view of our procedure
one could envisage the same procedure to get $R_q$ as for Shannon
entropy (with YES/NO questions). Both are maximal at equipartition
($p_i = 1/M$), and the maximum equals $\ln M$. The parameter $q$
of this entropy starts to act when distribution under
consideration is not uniform, otherwise $R_q$ is identical with
Shannon entropy (for $M$ cells one has $S=R_q=\ln M$ , whereas for
Tsallis entropy it is $S_q = \ln_q M$ ). Tsallis and Renyi
entropies are connected by (using the same $q$) $S_q =
\ln_q\left[\exp\left(R_q\right)\right]$.

\vspace{1cm}

\noindent The final version of this work owes much to discussions
at the {\it Facets of Entropy} workshop in Copenhagen (2007),
which GW gratefully acknowledges. Partial support (GW) of the
Ministry of Science and Higher
Education under contracts 1P03B02230 and  CERN/88/2006 is acknowledged.\\

\end{document}